\documentclass[a4paper,11pt]{article}
\pdfoutput=1 

\usepackage{jcappub} 

\usepackage[T1]{fontenc} 
\usepackage{amssymb}
\usepackage{float}
\usepackage{setspace}
\usepackage{verbatim}
\usepackage{graphicx}
\usepackage{dcolumn}
\newcommand{\bef}{\begin{figure}}
\newcommand{\eef}{\end{figure}}
\newcommand{\bec}{\begin{center}}
\newcommand{\eec}{\end{center}}
\newcommand{\be}{\begin{equation}}
\newcommand{\ee}{\end{equation}}
\newcommand{\rf}[1]{(\ref{eq:#1})}
\newcommand{\X}{\bar{X}}

\newcommand{\G}{{\cal G}}

\title{\boldmath On the thermal stability of a static spherically symmetric black holes in Nash embedding framework}

\author[a,b,1]{Abraao J. S. Capistrano,\note{Corresponding author.}}
\author[c,d]{Antonio C. Guti\'{e}rrez-Pi\~{n}eres,}
\author[e]{Sergio C. Ulhoa,}
\author[f]{Ronni G. G. Amorim.}

\affiliation[a]{Federal University of Latin-American Integration, 85867-970, P.o.b: 2123, Foz do Igua\c{c}u-PR, Brazil}
\affiliation[b]{Casimiro Montenegro Filho Astronomy Center, Itaipu Technological Park, 85867-900, Foz do Igua\c{c}u-PR, Brazil}
\affiliation[c]{Universidad Nacional Aut\'{o}noma de Mexico, Facultad de Ciencias Basicas, Universidad Tecnol\'{o}gica de Bol\'{i}var, Cartegena 13001, Colombia}
\affiliation[d]{Instituto  de  Ciencias  Nucleares,  Universidad  Nacional  Aut\'{o}noma  de  M\'{e}xico, AP  70543,  M\'{e}xico,  DF 04510,  M\'{e}xico}
\affiliation[e]{Instituto de F\'{i}sica, Universidade de Bras\'{i}lia, 70910-900, Bras\'{i}lia, DF, Brazil}
\affiliation[f]{Faculdade Gama, Universidade de Bras\'{i}lia, Setor Leste (Gama), 72444-240, Bras\'{i}lia, DF, Brazil}

\emailAdd{abraao.capistrano@unila.edu.br}
\emailAdd{acgutierrez@correo.nucleares.unam.mx}
\emailAdd{sc.ulhoa@gmail.com}
\emailAdd{ronniamorim@gmail.com}

\abstract{We study the deformation caused by the influence of extrinsic curvature on a vacuum spherically symmetric metric embedded in a five-dimensional bulk. In this sense, we investigate the produced black-holes and derive general characteristics such as their masses, horizons, singularities and thermal properties. As a test, we also study the bending of light near such black-holes analyzing the movement of a test particle and the modification caused by extrinsic curvature on its movement. Accordingly, using the asymptotically conformal flat condition for the extrinsic curvature, an analytical expansion of a set of \emph{n}-scalar fields can be defined and we show that the corresponding black holes must be large and constrained in the range of allowed values $-1/2 \leq n \leq 1.8$. As a result, they are locally thermodynamically stable, but not globally preferred.}


\begin{document}
\maketitle
\flushbottom

\section{Introduction}

In a previous publication \cite{cap2014}, we studied the influence of extrinsic curvature on the rotation curves of galaxies obtaining a good agreement with the observed velocity rotation curves of smooth hybrid alpha-HI measurements. In doing so, we used the smooth deformation concept \cite{cap2014,GDE,gde2,QBW} based on Nash theorem of embedding geometries \cite{Nash}. This allowed us the possibility to develop a more general approach to embedded space-times. Instead of assuming the string inspired embedding of a three-dimensional hypersurface generating a four-dimensional embedded volume, we look at the conditions for the existence of the embedding of the space-time. The embedding of a manifold into another is a non-trivial problem and the resulting embedded geometry resulting from of an evolving three-surface must comply the Gauss-Codazzi-Ricci equations. For instance, many of the currently brane-world models fail to comply with those equations.

In the Randall-Sundrum brane-world model (RS) \cite{RS,RS1}, the space-time is embedded in a five dimensional anti-deSitter space AdS5. It is assumed that the space-time acts as a boundary for the higher-dimensional gravitational field. This condition has the effect that extrinsic curvature becomes an algebraic function of the energy-momentum tensor of matter confined to the four-dimensional embedded space-time. In more than five dimensions that boundary condition does not make sense because the extrinsic curvature acquire an internal index while the energy-momentum tensor does not have this degree of freedom. To obtain the so-called Israel condition it becomes necessary to consider the brane-world as a boundary separating two regions of the bulk, and find the difference between the expression calculated in both sides. The Israel condition only follows after the additional condition that the boundary brane-world acts as a mirror: Only in this case the tangent components cancel and the components depending on extrinsic curvature remain. This can be seen in the original paper by W. Israel \cite{Israel}, or in the appendix B of reference \cite{GDE}.

We should add that the Israel condition is not used in many other brane-world models, such as, for instance, the original paper by the Arkani-Hamed, G. Dvali and S. Dimopolous \cite{add}, sometimes referred as ADD model, and  also the Dvali-Gabedadzi-Porrati model \cite{dgp}. Rather, it was shown that the Israel condition is not unique, so different conditions may lead to different physical results \cite{Brandon}. Other brane-world models have been developed with no need of particular junction conditions, see \cite{cap2014,GDE,gde2,QBW,maia2,sepangi,sepangi1,sepangi2} and references therein. Moreover, we should also add that such condition is particular to five-dimensional bulk spaces and is not consistent with higher dimensional embeddings. In the case of a Schwarzschild solution \cite{kasner,kasner2}, for instance, the embedding would be compromised in RS scheme and evinces the necessity of a more general framework. Unless we are restricted to a class particular models the Israel condition does not apply because only the mirror symmetry on the brane-world is dropped.

In principle general relativity is exempt of the discussed Riemann curvature ambiguity problem because the
ground state of the gravitational field, the Minkowski space-time, is well defined by the existence of Poincar\'{e}
symmetry (in special the translations). Such situation between particle physics and Einstein's
gravity was shaken by the observations of a small cosmological constant, then the emergence of the cosmological
constant problem. This leads us back to the necessity of the embedding of space-times as the only known solution of the Riemann
curvature ambiguity.

The general solution of the embedding problem (mainly on Gauss-Codazzi-Ricci equations) is shown in the paper by J. Nash on differentiable embedding \cite{Nash}, and generalized by Greene \cite{Greene} including positive and  negative  signatures. The application of Nash theorem to gravitation is a landmark of our paper, providing a new tool to the gravitational perturbation issue. The embedding mechanics presented here is a legitimate condition in all other more general settings.

The aim of our present work is to focus on the classical thermal stability of black holes embedded in five-dimensional bulk space-time. Recently, the study of the high dimensional space-times and the prediction of a new black objects (black holes, black strings, black rings, and so on) have been the focus of very active research \cite{emparam,albo,albo2,figueras}. Specifically, it is worth noting that a Schwarzschild embedding was a problem discussed a long ago \cite{kasner,kasner2,fronsdal,kruskal, Szekeres} and it is known that this geometry is completely embedded in six-dimensions \cite{monte}. In particular, the effects of a five dimensional bulk space-time endowed with an embedded spherically symmetric metric is still an open research arena since we do not have a closed analytical solution \cite{abdolrahimi, Harmark, Gorbonos, Gorbonos2}. Notwithstanding, it may lead to new physical consequences in a search of a full description and understanding of astrophysical objects, mainly on stellar dynamics and gravitational collapse. Moreover, in this work we explore the consequences of such embedding by using the asymptotic condition for extrinsic curvature in order to attenuate such restrictions.

The paper is organized as follows. In section 2, we briefly present a summary of the theoretical background of this paper restricted to the five dimensional space-time. In section 3, we study a spherically symmetric metric embedded in five dimensions. Moreover, in section 4, we analyze the constrains on the solution and the resulting horizons, which will be particular important to the thermodynamic analysis as presented in section 5. Finally, in the conclusion section, the final remarks and future prospects are presented.

\section{The induced four dimensional equations}
The present paper is not M-theory/Strings related. As well-known, the Randall-Sundrum (RS) model \cite{RS,RS1} is a brane-world theory originated from M-theory in connection with the derivation of the Horava-Witten heterotic $E8\times E8$ string theory in the space $AdS5\times S5$. In that framework, one obtains a compactification of one extra dimension on the orbifold $S1/Z2$ by using the $Z2$ symmetry on the circle $S1$.  The presence of the five dimensional anti-deSitter AdS5 space is mainly motivated by the prospects of the AdS/CFT correspondence between the superconformal Yang-Mills theory in four dimensions and the anti-deSitter gravity in five dimensions.

In a different approach, our proposal is initially based on the possibly to explore embedded space-times in a search of obtaining a more general theory based on embeddings (the brane-world models, of course, can be an example). Instead of assuming the string inspired embedding of a three-dimensional
hypersurface generating a four-dimensional embedded volume, we look at the conditions for the embedding of the space-time itself.

Nash's original  theorem   used  a  flat D-dimensional  Euclidean space  but this  was   soon generalized  to  any  Riemannian
manifold, including those  with   non-positive  signatures \cite{Greene}. For simplicity, we refer the term ``Nash theorem'' as valid also for pseudo-Riemannian manifolds. Another important aspect is that the seminal work on perturbation of geometry was proposed posthumously in 1926 by Campbell \cite{campbell} using analytic embedding functions and later extended to non-positive signatures \cite{romero}.
Although the Nash theorem  could  also be generalized to include  perturbations  on  arbitrary  directions in the bulk,    it
would  make  its interpretations  more difficult,   so that we
retain  Nash's   choice of  independent orthogonal  perturbations.
It  should be noted that  the   smoothness of the embedding is
a    primary  concern   of Nash theorem. In this  respect, the
natural  choice  for  the  bulk is that  its  metric  satisfy the
Einstein-Hilbert  principle. Indeed,  that principle represents  a
statement  on   the  smoothness  of the embedding  space (the variation  of the Ricci scalar  is the  minimum possible). Admitting that the perturbations are  smooth (differentiable),  then the  embedded  geometry  will  be  also differentiable.

As it happens, Nash showed that any embedded unperturbed metric $g_{\mu\nu}$ can be generated by a
continuous sequence of small metric perturbations of a given
geometry with a metric of the embedded space-time
\begin{equation}\label{eq:York}
\tilde{g}_{\mu\nu}  =  g_{\mu\nu}  +  \delta y^a \, k_{\mu\nu  a}  +
\delta y^a\delta y^b\, g^{\rho\sigma}
k_{\mu\rho a}k_{\nu\sigma b}+\cdots\;.
\end{equation}
or, equivalently
\begin{equation}\label{eq:York2}
k_{\mu\nu}=-\frac{1}{2}\frac{\partial g_{\mu\nu}}{\partial y} \;,
\end{equation}
where $k_{\mu\nu}$ is the initially unperturbed extrinsic curvature and  $y$ is the coordinate related to extra dimensions. Since Nash's smooth deformations are applied to the embedding process, the coordinate $y$, usually noticed in ridig embedded models, e.g \cite{RS,RS1}, can be omitted in the process for perturbing an element line (depending on what one wants to do there are many different ways to embed a manifold into another classified as local, global, isometric, conformal or more generally defined by a collineation, rigid, deformable, analytic or differentiable embedding). Differently from RS models, using Nash embedding framework, we can develop a model that considers the extrinsic curvature as a dynamical (physical) quantity \cite{cap2014,GDE,gde2,QBW,maia2,sepangi,sepangi1,sepangi2}. Also, the paper should not be confused with a paper on quantization deformation, since our arguments are all classical.

To start with, we consider  an example with a  Riemannian  manifold  $\bar{V_4}$ with  metric $\bar{g}_{\mu\nu}$,  and its  local  isometric  embedding in a   five-dimensional  Riemannian  manifold   $V_5$, given  by  a   differentiable  and  regular map $\bar{X}: \bar{V}_4 \rightarrow  V_5$  satisfying the  embedding  equations
\be
\X^A{}_{,\mu} \X^B{}_{,\nu}\G_{AB}=g_{\mu\nu},\;  \X^A{}_{,\mu}\bar{\eta}^B \G_{AB}=0,  \;  \bar{\eta}^A \bar{\eta}^B \label{eq:X}\G_{AB}=1,\;
A,B = 1..D
\ee
where  we have  denoted by   $\G_{AB}$  the metric components of  $V_5$  in  arbitrary  coordinates,  and where  $\bar{\eta}$  denotes  the unit  vector  field   orthogonal  to  $\bar{V}_4$. Concerning notation, capital Latin indices run from 1 to 5. Small case Latin indices refer to the only one extra dimension considered. All Greek indices refer to the embedded space-time counting from 1 to 4. In general we have $D= n+ 1$ with the index $D$ representing the bulk space dimension and the index $n$ represents the embedded space dimension.

 The   extrinsic  curvature of  $\bar{V}_n$  is by  definition  the   projection of  the  variation of $\eta$   on the tangent plane :
\be
\bar{k}{}_{\mu\nu} =  -\bar{X}^A{}_{,\mu}\bar{\eta}^B{}_{,\nu}  \G_{AB}=\X^A{}_{,\mu\nu}\bar{\eta}^B \G_{AB} \label{eq:extrinsic}
\ee
The  integration of the   system of  equations  Eq.\rf{X}   gives  the   required   embedding  map  $\bar{X}$. Moreover, one can construct   the    one-parameter  group of  diffeomorphisms  defined   by the  map   $h_{y}(p): V_D\rightarrow V_D$,  describing  a  continuous curve    $\alpha(y)=h_y (p)$, passing  through  the  point  $p \in \bar{V_n}$,  with  unit    normal  vector  $\alpha'(p) =\eta(p)$. Thus, the   group  is  characterized by   the composition  $h_{y} \circ h_{\pm y'}(p)\stackrel{def}{=} h_{y \pm y'}(p)$,  $h_{0}(p)\stackrel{def}{=}p$. Accordingly, applying   this  diffeomorphism   to   all points of  a small  neighborhood of  $p$,  we  obtain  a   congruence   of  curves (or orbits)   orthogonal  to   $\bar{V}_n$. It   does  not matter  if   the parameter  $y$  is  time-like or  not,   nor  if  it is  positive or  negative.

If one defines  a  geometric  object $\bar{\omega}$   in  $\bar{V}_n$,  its Lie  transport along the  flow   for  a small distance  $\delta y$    is  given  by     $\Omega   = \bar{\Omega}  + \delta y  \pounds_\eta{\bar{\Omega}}$,  where $\pounds_\eta$  denotes the Lie  derivative  with respect  to   $\eta$.   In particular,   the  Lie  transport of the  Gaussian frame     $\{\X^A_\mu ,  \bar{\eta}^A_a  \}$, defined on  $\bar{V}_n$   gives
\begin{eqnarray}
Z^A{}_{,\mu}  &=&  X^A{}_{,\mu}  +   \delta y \;\pounds_\eta{X^A{}_{,\mu}}
=  X^A{}_{,\mu} + \delta y \;  \eta^A{}_{,\mu}\label{eq:pertu1}\\
  \eta^A  &=&\bar{\eta}^A  +   \delta y\;[\bar{\eta}, \bar{\eta}]^A
\;\;\;\;\;\;= \;\;\bar{\eta}^A \label{eq:pertu2}
\end{eqnarray}
However,  from  Eq.\rf{extrinsic}  we note  that  in general   $\eta_{,\mu}  \neq \bar{\eta}_{,\mu}$.

In order  to describe  another  manifold, the  set of  coordinates  $Z^A$ need  to  satisfy the embedding  equations similar  to  Eq.\rf{X} as
\be
Z^A{}_{,\mu} Z^B{}_{,\nu}\G_{AB}=g_{\mu\nu},\;  Z^A{}_{,\mu}\eta^B \G_{AB}=0,  \;  \eta^A \eta^B \G_{AB}=1 \label{eq:Z}
\ee
 Replacing   Eq.\rf{pertu1} and   Eq.\rf{pertu2} in  Eq.\rf{Z} and  using  the  definition from Eq.\rf{extrinsic}, we obtain  the    metric  and  extrinsic  curvature of  the   new  manifold
\begin{eqnarray}
&&g_{\mu\nu} =   \bar{g}_{\mu\nu}-2y \bar{k}_{\mu\nu} + y^2 \bar{g}^{\rho\sigma}\bar{k}_{\mu\rho}\bar{k}_{\nu\sigma}\label{eq:g}\\
&&k_{\mu\nu}  =\bar{k}_{\mu\nu}  -2y \bar{g}^{\rho\sigma} \bar{k}_{\mu\rho}\bar{k}_{\nu\sigma}  \label{eq:k1}
\end{eqnarray}
Taking  the  derivative  of  Eq.\rf{g} with  respect  to  $y$ we  obtain  Nash's  deformation   condition  as shown in Eq.(\ref{eq:York2}). Moreover, the  integrability  conditions  for   these  equations   are   given  by  the  non-trivial  components of  the  Riemann  tensor of  the embedding  space\footnote{To  avoid  confusion  with  the  four dimensional  Riemann tensor $R_{\alpha\beta\gamma\delta}$, expressed    in the frame
$\{ Z^A_\mu, \eta^A  \}$  as
\begin{eqnarray}
&&\hspace{-7mm}^5{\cal R}_{ABCD}Z^A{}_{,\alpha}Z^B{}_{,\beta}Z^C{}_{,\gamma}Z^D{}_{,\delta} = R_{\alpha\beta\gamma\delta} +\!\!
(k_{\alpha\gamma}k_{\beta\delta}\!-\!
k_{\alpha\delta}k_{\beta\gamma})\!\!
\label{eq:G1}\\
&&\hspace{-7mm}^5{\cal R}_{ABCD}Z^A{}_{,\alpha} Z^B{}_{,\beta}Z^C{}_{,\gamma}\eta^D=k_{\alpha[\beta;\gamma]}  \label{eq:C1}
\end{eqnarray}
These   are  the   mentioned   Gauss-Codazzi  equations
(the  third  equation -the Ricci  equation- does  not appear  in the  case of  just one  extra  dimension).  The  first  of these  equation (Gauss) shows  that the  Riemann   curvature of  the  embedding  space  acts  as  a  reference  for  the  Riemann  curvature  of  the   embedded  space-time.
The  second  equation (Codazzi) complements  this  interpretation,  stating  that projection of  the  Riemann  tensor  of  the embedding space  along the  normal  direction  is  given  by  the tangent variation of  the  extrinsic  curvature. Notice  from   Eq.\rf{G1} that  the   local  shape of  an  embedded  Riemannian  manifold  is  determined  not only  by  its  Riemann  tensor  but  also  by  its  extrinsic  curvature,  completing  the  proof  of  the  Schlaefli  embedding  conjecture  by  use  of  Nash's   deformation condition  in Eq.(\ref{eq:York2}). This guarantees to reconstruct the five-dimensional geometry and understand its properties from the dynamics, in this case, of the four-dimensional embedded space-time.}.

As in Kaluza-Klein   and   in the brane-world  theories,  the  embedding  space   $V_5$  has  a  metric  geometry  defined   by  the  higher-dimensional Einstein's  equations for the  bulk in  arbitrary  coordinates
 \be
{\mathcal{R}}_{AB}-\frac{1}{2}\mathcal{R} \mathcal{G}_{AB}
 =\alpha_* T^*_{AB}\;,\label{eq:EEbulk}
\ee
 where the metric of the bulk is denoted by  ${\cal G}_{AB}$ and we have dispensed the bulk with a cosmological constant, since for the present application in astrophysical scale the induced four-dimensional cosmological constant has a very small value and can be neglected. The energy-momentum tensor for the bulk of the known  matter and gauge fields is denoted by $T^*_{AB}$. The  constant  $\alpha_*$ determines, in this case, the five-dimensional energy scale .

In what concerns the confinement, the  four-dimensionality of the    space-time is  an
experimentally  established fact,  associated  with  the   Poincar\'{e}
invariance of  Maxwell's  equations and  their  dualities, later
extended to all gauge fields. Therefore,   all matter  which
interacts   with these gauge  fields  must be defined in  the  four-dimensional  space-times.  On the other
hand,  in  spite of  all  efforts  made so far, the gravitational
interaction  has failed to  fit into  a  similar  gauge  scheme,  so
that the  gravitational  field  does  not necessarily have the same
four-dimensional  limitations and accesses  the extra
dimensions in accordance  with Eq.\rf{York},  regardless  the  location
of  its  sources.

We  assume  that the  four-dimensionality   of   gauge   fields  and  ordinary matter  applies  to all perturbed  space-times,  so  that  it  corresponds  to   a  confinement  condition.  In order  to   recover  Einstein's gravity  by reversing the  embedding, the   confinement  of  ordinary matter and  gauge  fields implies that the  tangent  components of  $\alpha_* T^*_{AB}$  in Eq.(\ref{eq:EEbulk}) must  coincide  with ( $8\pi G  T_{\mu\nu}$), where
$T_{\mu\nu}$  is the  induced four-dimensional energy-momentum  tensor of the confined
sources. As it may  have  been already noted,   we  are
essentially  reproducing the  brane-world  program,  with the
difference that  it is  a  general approach and  has  nothing to  do
with branes in string/M theory. Instead,  all  that we  use here is
Nash theorem  together  with  the  four-dimensionality of gauge
fields,  the  Einstein-Hilbert principle for the bulk and  a
D-dimensional energy scale $\alpha_*$. The  confinement  is set simply as $\alpha_* T^* =  8\pi G
 T_{\mu\nu}$, where $T_{\mu\nu}$ represents the energy momentum
tensor  of the confined  matter, and  $T^*_{\mu 5}=T^*_{55}=0$. This can be understood in more general terms as a consequence of the duality operations of the Yang-Mills equations. It follows that $D\wedge F^{\ast}$ is a three-form that must be isomorphic to the current 1-form $j^{\ast}$. This condition
depends only on the four-dimensionality of the space-time \cite{Donaldson, Taubes} consistent with a plethora of experimental facts. Even that any gauge theory can be mathematically constructed in a higher dimensional space, we adopt the confinement as a condition, since the four-dimensionality of space-time will suffice in our case based on experimentally high-energy tests suggest \cite{lim}.

With  the  above  remarks,  we may   re-write the  components   of  Eq.(\ref{eq:EEbulk}), using Eq.(\ref{eq:York2}) and Eq.(\rf{Z}) together  with the previous confinement conditions \cite{GDE}  obtaining
\begin{eqnarray}\label{eq:BE1}
&&R_{\mu\nu}-\frac{1}{2}Rg_{\mu\nu}-Q_{\mu\nu}=
8\pi G T_{\mu\nu}\; \hspace{2mm}\\
&&\label{eq:BE2} k_{\mu;\rho}^{\;\rho}-h_{,\mu} =0\;,\hspace{4,9cm}
\end{eqnarray}
where  the  term   $Q_{\mu\nu}$ results  from the  expression  of  $\mathcal{R}_{AB}$ in Eq.(\ref{eq:EEbulk}),  involving  the  orthogonal  and  mixed    components  of  the  Christoffel  symbols
\begin{equation}\label{eq:qmunu}
Q_{\mu\nu}=g^{\rho\sigma}k_{\mu\rho }k_{\nu\sigma}- k_{\mu\nu }h -\frac{1}{2}\left(K^2-h^2\right)g_{\mu\nu}\;,
\end{equation}
where     $h^2= g^{\mu\nu}k_{\mu\nu}$ is  the  (squared) mean curvature and
 $K^{2}=k^{\mu\nu}k_{\mu\nu}$  is  the  (squared)  Gauss curvature.
This   quantity   is  therefore   entirely  geometrical   and  it  is  conserved in the sense of
\begin{equation}\label{eq:cons}
Q^{\mu\nu}{}_{;\nu}=0\;.
\end{equation}
Therefore  we  may    derive  observables effects associated  with the  extrinsic  curvature   capable  to be seen  by
four-dimensional observers  in space-times. A detailed derivation of these equations can be found in \cite{GDE,gde2,QBW} and references therein. Hereafter, we use a system of unit such that $c=G=1$.

\section{Induced spherically symmetric vacuum solution}
In the present study, we focus our work on obtaining an exterior vacuum spherical solution, i.e, $T_{\mu\nu}=0$. To this end, we start with the general static spherically symmetric induced metric that can be described by the line element
\begin{equation}\label{eq:line element}
ds^2 = B(r) dt^2 -  A(r) dr^2 - r^2 d\theta^2 - r^2 \sin^2\theta d\phi^2\;\;,
\end{equation}
where we denote the functions $A(r)=A$ and $B(r)=B$. Thus, one can obtain the following components for the Ricci tensor:
$$R_{rr} = \frac{B''}{2B} - \frac{1}{4}\frac{B'}{B} \left(\frac{A'}{A} + \frac{B'}{B} \right) - \frac{1}{r} \frac{A'}{A}$$
$$R_{\theta\theta} = -1 + \frac{r}{2A} \left(-\frac{A'}{A} + \frac{B'}{B} \right) + \frac{1}{A}$$
$$R_{\phi\phi} = \sin^2\theta R_{\theta\theta}$$
$$R_{tt} = -\frac{B''}{2A} + \frac{1}{4}\frac{B'}{A} \left(\frac{A'}{A} + \frac{B'}{B} \right) - \frac{1}{r} \frac{B'}{A}$$
where we have $\frac{dA}{dr} = A'$ e $\frac{dB}{dr} = B'$.

From Eq.(\ref{eq:BE1}), the gravitational-tensor vacuum equations (with $T_{\mu\nu}=0$) can be written in alternative form as
\begin{equation}\label{eq:1}
    R_{\mu\nu} + \frac{1}{2} Q g_{\mu\nu} = Q_{\mu\nu}
\end{equation}
where we use the contraction $Q= g^{\mu\nu} Q_{\mu\nu}$.

The general solution of Codazzi equations Eq.(\ref{eq:BE2}) is given by
\begin{equation}\label{eq:geneqk}
    k_{\mu\nu}=f_{\mu}g_{\mu\nu} \;\;\;\;(no\;sum\;on\;\mu)\;,
\end{equation}
Taking the former equation and the definition of $Q_{\mu\nu}$, one can write
$$Q_{\mu\nu}= f^2_{\mu}g_{\mu\nu}-\sum_{\alpha}f_{\alpha}f_{\mu}g_{\mu\nu}-\frac{1}{2}\left(\sum_{\alpha}f^2_{\alpha}
-\left(\sum_{\alpha}f_{\alpha}\right)^2\right)g_{\mu\nu}\;,$$
where
$$U_{\mu}=f^2_{\mu}-\left( \sum_{\alpha}f_{\alpha}\right)f_{\mu}
-\frac{1}{2}\left(\sum_{\alpha}f^2_{\alpha}-\left(\sum_{\alpha}f_{\alpha}\right)^2\right)\delta^\mu_\mu\;.$$
Consequently, we can write $Q_{\mu\nu}$ in terms of $f_{\mu}$ as
\begin{equation}
Q_{\mu\nu}= U_{\mu}g_{\mu\nu} \;\;\;\;\;(no\;\;sum\;\;on\;\;\mu).
\end{equation}
Since $Q_{\mu\nu}$ is a conserved quantity, we can find four equations from $\sum_{\nu} g^{\mu\nu}U_{\mu;\nu}=0$ that can be reduced to only two equations:
$$
\left\{
  \begin{array}{ll}
    \left(f_1 + f_2\right) \left(f_3 - f_4\right) = 0 & \hbox{;} \\
    \left(f_3 + f_4\right) \left(f_1 - f_2\right) = 0 & \hbox{.}
  \end{array}
\right.
$$
and result in the condition
\begin{equation}\label{eq:conditio}
f_1 f_3 = f_2 f_4\;.
\end{equation}

A straightforward consequence of the homogeneity of Eq.(\ref{eq:BE2}) and the condition Eq.(\ref{eq:conditio}), the individual arbitrariness of the functions $f_{\mu}$ can be reduced to a unique arbitrary function $\alpha$ that depends on the radial coordinate. Hence, the equation (\ref{eq:geneqk}) turns out to be
\begin{equation}\label{eq:alphak}
 k_{\mu\nu} = \alpha(r) g_{\mu\nu}\;,
\end{equation}
with $\alpha(r)=\alpha$.

Thus, we can determine the extrinsic quantities
\begin{eqnarray}
   Q_{\mu\nu} & = &  3 \alpha^2 g_{\mu\nu}\;,\\
   Q = Tr(Q_{\mu\nu})& = &  12 \alpha^2\;,
\end{eqnarray}
and substituting in the gravitational-tensor equation Eq.(\ref{eq:BE1}), one can obtain the following $rr$ and $tt$ components:
\begin{eqnarray}
\frac{B''}{2B} - \frac{1}{4}\frac{B'}{B} \left(\frac{A'}{A} + \frac{B'}{B} \right) - \frac{1}{r} \frac{A'}{A}= 9 \alpha^2(r)A\;,\\
-\frac{B''}{2A} + \frac{1}{4}\frac{B'}{A} \left(\frac{A'}{A} + \frac{B'}{B} \right) - \frac{1}{r} \frac{B'}{A}= -9\alpha^2(r)B\;.
\end{eqnarray}
From these equations, we have
$$\frac{A'}{A} =- \frac{B'}{B}\;.$$
thus, $AB =\;constant\;.$

In the same way as the calculation of the very known Schwarzschild vacuum solution, we impose the contour
$$\lim_{r \rightarrow \infty }A(r) = \lim_{r \rightarrow \infty}  B(r)= 1\;,$$
in order to approach the metric tensor to Minkowski tensor as $r \rightarrow \infty$, we have
$$A(r) = \frac{1}{B(r)}\;.$$
Furthermore, using the $(\theta, \theta)$ component, one can obtain
\begin{equation}\label{eq:br1}
B(r) = 1 + \frac{k}{r} + \frac{9}{r} \int \alpha^2(r) r^2 dr\;,\\
\end{equation}
and
\begin{equation}\label{eq: ar}
A(r)= \left[B(r)\right]^{-1} = \left[ 1 + \frac{k}{r} + \frac{9}{r} \int \alpha^2(r) r^2 dr\; \right]^{-1}\;.
\end{equation}
where $k$ is a constant.

Since spherically symmetric geometry is very constrained in a five-dimensional embedding \cite{monte}. Based on the behavior of extrinsic curvature at infinity, we set the asymptotically conformal flat condition on extrinsic curvature as
\begin{equation}\label{eq:flatcondition}
\lim_{r\rightarrow \infty} k_{\mu\nu} = \lim_{r\rightarrow \infty} \alpha(r)\lim_{r\rightarrow \infty} g_{\mu\nu}\;.
\end{equation}
It is important to stress that the Nash theorem \emph{per se} does not provide a dynamical set of equations for extrinsic curvature but shows how to relate extrinsic curvature to the metric through a smooth process of deformation making the local embedding regular and completely differentiable. As $\lim_{r\rightarrow \infty} g_{\mu\nu} \rightarrow \eta_{\mu\nu}$, where $\eta_{\mu\nu}$ is Minkowskian $M_{4}$ metric, the extrinsic curvature vanishes as it tends to infinity, so the function $\alpha(r)$ must comply with this condition. Thus, we can infer that the function $\alpha(r)$ must be analytical at infinity and to attend this constraint, we choose the simplest option
\begin{equation}\label{eq:alfaform}
\alpha(r)= \sum_{n=i}^{s}\frac{\sqrt{-\alpha_0}}{\gamma^{*} r^n}\;,
\end{equation}
where the sum is upon all scalar potentials and the indices $i$ and $s$ are real numbers. The index $n$ may represent a set of scalar fields. The parameter $\alpha_0$ has cosmological magnitude with the same units as the Hubble constant and its modulus is estimated as $0.677$ \cite{cap2014} that will be used hereon. In order to keep the right dimension to Eq.(\ref{eq:alfaform}), we have  introduced a unitary parameter $\gamma^{*}$ that has the inverse unit of Hubble constant \cite{cap2014} and defines the cosmological horizon in Eq.(\ref{eq:line element1}). From the geometrical point of view, Eq.(\ref{eq:alfaform}) will not produce umbilical points on this spherically surface leading to a local additional bending in space-time.

Using Eqs.(\ref{eq:br1}) and (\ref{eq:alfaform}), one can obtain an explicit form of the coefficient $B(r)$  given by
\begin{equation}\label{eq:br2}
B(r) = 1 - \frac{K(9\alpha^2_0 + 1)}{r} - \sum_{n=i}^{s}\frac{9\alpha_0}{\gamma^{*}(2n-3)} r^{2\left(1-n\right)}\;.\\
\end{equation}
In terms on the correspondence principle with Einstein gravity, we set $K(9\alpha_0 + 1)=-4M$, which remains valid even in the limit when $\alpha\rightarrow 0$ in order to obtain the asymptotically flat solution. In addition, the related potential is given by
\begin{equation}\label{eq:potential}
\Phi(r) = - 1 + \frac{2M}{r} + \sum_{n=i}^{s}\frac{9\alpha_0}{\gamma^{*}(3-2n)} r^{2\left(1-n\right)}\;,\\
\end{equation}
evincing how the initially spherical symmetric geometry is modified by the influence of the extrinsic terms.
Thus, we can write the line element in Eq.(\ref{eq:line element}) modified by extrinsic curvature as
\begin{equation}\label{eq:line element1}
  ds^2 =   \Phi(r) dt^2   -  \frac{1}{\Phi(r)} dr^2 - r^2 d\theta^2 - r^2 \sin^2\theta d\phi^2\;,
\end{equation}
for all $n\neq \frac{3}{2}$. Hence, one can obtain physical singularities with the calculation of the induced four-dimensional Kretschmann scalar $\tilde{K}$ and find
$$\tilde{K}=\frac{ g(r,\alpha_0/\gamma^{*},M,n)}{r^6{(3-2n)}^2 ((r-2M)(3-2n)r^{2n}+9(\alpha_0/\gamma^{*})^2r^3)^6}\;,$$
which preserves at first an intrinsic, physical and irremovable singularity at $r=0$ and $n=\frac{3}{2}$.

Another situations can be found when considering the term $g(r,\alpha_0,M,n)$. This term represents a very large polynomial function with 384 terms that were omitted here in its explicit form. Bearing in mind that the value of the extrinsic parameter $\alpha_0$ has magnitude $0.677$ or can be zero (if considering a vanishing extrinsic curvature) and analyzing the behavior of $g(r,\alpha_0,M,n)$, one can summarize in the following table other singularities, a vanishing and positive $\tilde{K}$.

As we can see, from the second to the fourth row in table (\ref{tab:kre}) for any different values of $n$, the term $g(r,\alpha_0,M,n)$  diverges leading to an undefined $\tilde{K}$. The fifth and sixth rows show an expected result since Riemannian curvature takes over when we have a vanishing extrinsic curvature. Interestingly, the last row shows the influence of the distortion cause by  extrinsic curvature since even with a (non-forming) black hole with zero-mass \cite{gib}, local tilde effects can be found once we do not have as a result the Minkowskian flat space-time. The range $-10\leq n \leq 10$ was found once a larger range induces a very fast growing and decaying of the exponents but not observed in solar scales. In the section 4, we will study the possibility to get a more constrained range for those values of $n$.
\begin{table*}
    \centering
     \begin{minipage}{100mm}
        \begin{tabular}{@{}ccccccccc@{}}
  \hline
  r    &     $\alpha_0$ &      n               &      Mass      &      $g(r,\alpha_0,M,n)$                & $\tilde{K}$ \\
  \hline
  0    &      0         & $0< n \leq 6$        &       any      &          0                              & undefined \\
  0    &      0         &      0               &       any      &          $>0$                           & diverges \\
  0    &    0.677       & $0\leq n \leq 6$     &       any      &         $ >0$                           & diverges \\
 $ >0$ &     0          &      any             &       0        &           0                             &   0       \\
 $ >0$ &     0          &      any             &       $ >0$    &           0                             &   $ >0$    \\
  $>0$ &    0.677       & $-10\leq n \leq 10$  &       $\geq 0$ &         $ >0 $                          &  $ >0$      \\
  \hline
\end{tabular}
    \caption{A summary of relevant values for the term $g(r,\alpha_0,M,n)$ in determining different behaviors of Kretschmann scalar $\tilde{K}$. }\label{tab:kre}
     \end{minipage}
\end{table*}

For the sake of completeness, we test the bending of light near the horizons in this geometry warped by Nash's embedding. Using the line element for spherical symmetry in Eq.(\ref{eq:line element}), one can study the geodesic motion given as the solution of the following equations
\begin{eqnarray}
\frac{d}{d\tau}(-A\dot{t})&=&0\,,\nonumber\\
\frac{d}{d\tau}(r^2\dot{\theta})-r^2\sin\theta\cos\theta\dot{\phi}^2&=&0\,,\nonumber\\
\frac{d}{d\tau}(r^2\sin^2\theta\dot{\phi})&=&0\nonumber\\
-A\dot{t}^2+A^{-1}\dot{r}^2+r^2\dot{\theta}^2+r^2\sin^2\theta\dot{\phi}^2&=&0\,,\nonumber\\
\end{eqnarray}
where $\tau$ is the proper time and $\frac{d x}{d\tau}\equiv\dot{x}$. In addition, considering that the movement of a test particle takes place on a plane $\theta=\pi/2$, we perform a change of variables $U=r^{-1}$ and find the following equation of trajectory
\begin{equation}
\frac{d^2U}{d\phi^2}+AU+\left(\frac{dA}{dU}\right)\left(\frac{U^2}{2}\right)=0\,,
\end{equation}
which yields
\begin{equation}
\frac{d^2U}{d\phi^2}+U=3MU + \sum_{n=i}^{s}\frac{9n(\alpha_0/\gamma^{*}) U^{2n-1}}{(3-2n)}\,.
\end{equation}
Moreover, one can obtain the deflection angle as given by
\begin{equation}
\Delta\phi=\sum_{n=i}^{s} \int_b^\infty\frac{2dr}{r\left[\left(\frac{r}{b}\right)^2-A\right]^{1/2}}-\pi\,,
\end{equation}
where the parameter \emph{b} is the minimal distance from the source. Therefore,
\begin{equation}
\Delta\phi=\frac{4M}{b}-\sum_{n=i}^{s} \int_b^\infty\frac{9\alpha_0/\gamma^{*}r^{1-2n}dr}
{(3-2n)\left[\left(\frac{r}{b}\right)^2-1\right]^{3/2}}\,.
\end{equation}
We note that
\begin{equation}
\int\frac{r^{1-2n}dr}
{\left[\left(\frac{r}{b}\right)^2-1\right]^{3/2}}=-\left(\frac{r^{2-2n}
{}_2F_1(3/2,1-n,2-n,r^2/b^2)}{2(1-n)}\right)\,,
\end{equation}
where ${}_2F_1$ is the gaussian hypergeometric function. Thus, one can find a generic expression for the shift $\Delta\phi$ as
\begin{equation}
\Delta\phi=\Delta\phi_{GR}+\Delta\phi_{d}\;,
\end{equation}
where $\Delta\phi_{GR} = \frac{4M}{b}$ is the classical result provided from general relativity, and the second term $\Delta\phi_{d}$ which is given by
\begin{equation}
\Delta\phi_{d}= \frac{9\alpha}{2\gamma^{*}}\sum_{n=i}^{s} \frac{r^{2(1-n)}}{(2n^2-5n+3)}\;\;\;{}_2F_1(3/2,1-n,2-n,r^2/b^2)\;,
\end{equation}
which measures how strong is the deviation that depends on how intense is the deformation on the local space-time. For the sake of completeness,  if we neglect $\alpha_0$, one obtains the same result obtained from general relativity.

\section{Constraints on the allowed values of the \emph{n}-scalar fields}
In order to study the possible horizons by the element line in Eq.(\ref{eq:line element1}), we must constrain the values of the set of \emph{n}-scalar fields. We start with analyzing the behavior of the gravitational potential on solar scales. In general, black holes are objects placed in galactic scales ruled essentially by newtonian potentials ($\sim\frac{1}{r}$) and its smooth deviations in decaying $(\sim\frac{1}{r^2},\sim\frac{1}{r^3})$ and growing not large as $\sim r^2$. The increasing of gravitational potentials of order of $\sim r^3$ possibly can lead to a topological deformation of the space-time \cite{visser} or quadratically, and so on, should be quite worrisome and inconsistent with the solar system scale that turns the possible values of $n$ dramatically constrained. Bearing this in mind, we analyze the allowed values of $n$, since we already have a first insight from the analysis of Kretschmann scalar in table (\ref{tab:kre}).

The general expression for the horizons of Eq.(\ref{eq:line element1}) can be found when we set $g_{tt}=0$, so one can find
\begin{equation}\label{eq: horizons}
\sum_{n=i}^{s}\frac{9\alpha_0}{3-2n} r^{3-2n} + r - 2M=0\;,\\
\end{equation}
which is a polynomial equation of order $(3-2n)$ that gives a class of horizons to be defined for all allowed values of $n$. Moreover, the corresponding mass is given by
\begin{equation}\label{eq:mass}
M= \frac{1}{2}\left[\sum_{n=i}^{s}\frac{9\alpha_0}{3-2n} r^{2n-3} + r \right] \;.\\
\end{equation}
In order to avoid overloaded notation, the unitary parameter $\gamma^{*}$ is dividing the parameter $\alpha_0$ hereon. Hence, we obtain interesting cases for the allowed values of $n$ since the value $n=3/2$ is discarded due to the fact that it diverges Eq.(\ref{eq:br2}).

For instance, the Schwarzschild horizon can be straightforwardly  obtained when we set $\alpha_0=0$, which means that extrinsic curvature vanishes and the space-time turns out to be asymptotically Minkowskian. For $n=0$, one can recover the Schwarzschild-de Sitter-like solution and $n=2$ mimics  Reissner-N\"{o}rdstrom solution. It is important to say that the Reissner-N\"{o}rdstrom-like solution the term $\alpha_0^2$ is interpreted as a tidal charge but not originated from a electromagnetic field, since it is a vacuum solution. Interestingly, in this particular case, differently from RS models, our tidal charge $\alpha_0$ is originated from the extrinsic curvature and has a new physics associated to it, since it has a cosmological magnitude. The solution when $n=1$ alone is a modification of the Schwarzschild solution by a term  $9\alpha_0$.

For $n\geq3$, the solutions produce a very low newtonian potentials ($\sim \frac{1}{r^4}$ and on). On the other hand, for $n \leq -1$ the potentials grows very fast. Both limits are not verified in solar system scales. Accordingly, we can set that the parameter $n$ in summation must vary such as $-1/2 < n < 3$ hereon.

An interesting case can be found when we set $n= 1/2$ where we have found the equation
$$r-2M-\alpha_0r^2= 0\;,$$
with cosmological horizon $r_c= \frac{1+\sqrt{1-8\alpha_0M}}{2\alpha_0}$ and black hole horizon $r_{bh}= \frac{1-\sqrt{1-8\alpha_0M}}{2\alpha_0}$. Moreover, they exist when $1-8\alpha_0M >0$ for $M=\frac{1}{8\alpha_0}$ and both horizons coalesce at $r_c=r_{bh}= \frac{1}{2\alpha_0}$. Interestingly, taking eq.(\ref{eq:br2}), one can set $n= \frac{3-\beta}{2}$, where $\beta$ is a parameter, and find the same power law found in \cite{capozz}, which was used for an alternative model to explanation of the dark matter problem. Another curious fact from the richness of this model is that when $n=5/2$, which mimics an asymptotically metric function for a Bardeen black hole \cite{bardeen}.

\section{Thermodynamical stability}
In order to study the classical stability of this model, we focus on the determination of heat capacity and the resulting free energy. To this end, we calculate the related Hawking temperature, which depends on how curved is the space near a black hole. Therefore, one can find the local surface gravity as
$$\kappa = -\frac{1}{2} \frac{d g_{44}}{dr}|_{r= r_h (\;horizon)} = \frac{M}{r_{h}}9\alpha_0^2 \frac{(1-n)}{(3-2n)} r_h^{1-2n}\;.$$
Consequently, the Hawking temperature is given by
\begin{equation}\label{eq:hwtemp}
T= \frac{\kappa}{2\pi}= \frac{1}{4\pi r_h^2}\left[r_h + 9\alpha_0^2 r_h^{3-2n}\right]\;,
\end{equation}
where we denote the event horizon $r_h$. For the sake of completeness, we point out that when $\alpha_0 \rightarrow 0$ we recover Schwarzschild with event horizon $r_h= 2M$ and Hawking temperature $T_h= \frac{1}{8\pi M}$.

In addition, we determine the entropy defined by Bekenstein-Hawking formulae as
$$
S= \int^{r_h}_{0} \frac{1}{T} \frac{\partial M}{\partial r} dr\;,
$$
and using the corresponding mass in Eq.(\ref{eq:mass}), one can find
\begin{equation}\label{eq:entropy}
S= \pi r_h^2\;,
\end{equation}
which is interesting since it preserves the positivity of the entropic growth.

\begin{figure*}
\centering
\includegraphics[width=6cm,height=6cm]{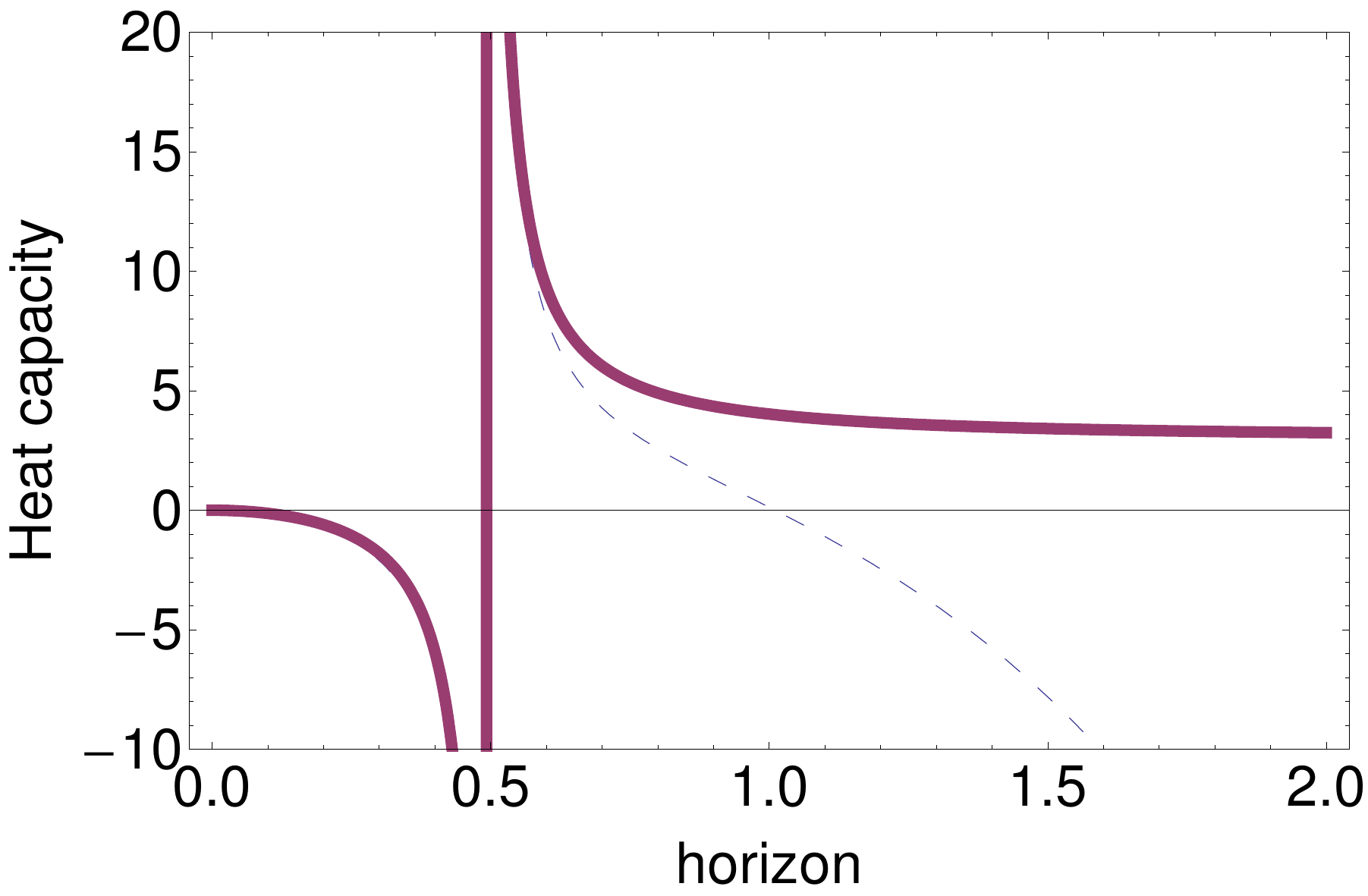}
\includegraphics[width=6cm,height=5.7cm]{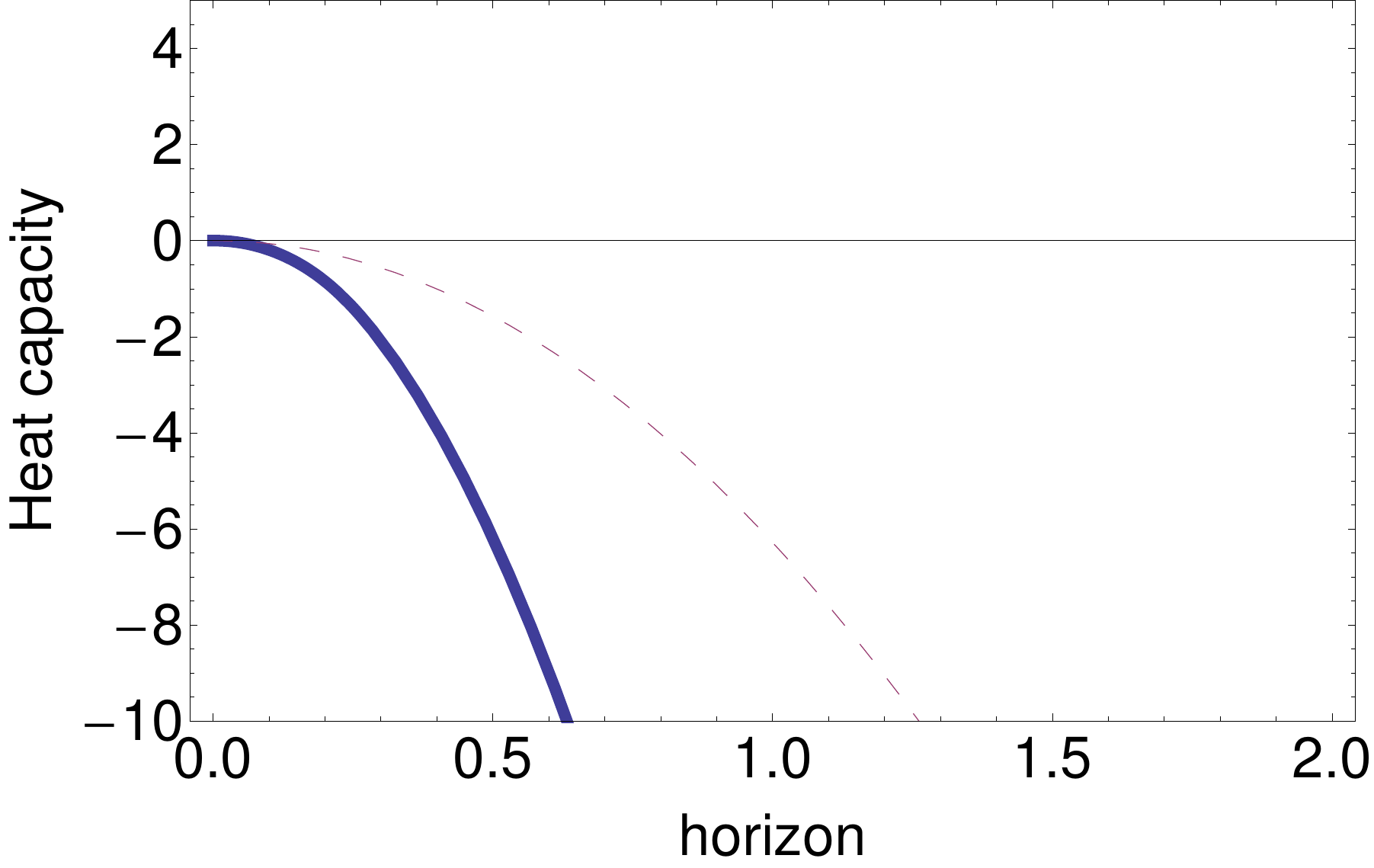}
\caption{The heat capacity is plotted as a function of the horizon with the allowed values of $n$. In the left panel is shown the behavior for the range in the allowed values $-1/2 \leq n \leq 1.8$ (solid line) and for $-1/2 \leq n \leq 3$ (dashed line). In the right panel, we compare the case $-1/2 \leq n \leq 3$ (solid line) to Schwarzschild solution (dashed line).}\label{fig:first}
\end{figure*}

From the classical expression of heat capacity, one can find the alternative form in terms of the mass of the black hole
$$C= \left(\frac{\partial M}{\partial T}\right) =\; \left(\frac{\partial M}{\partial r_h}\right)\;\left(\frac{\partial r_h}{\partial T}\right)\;,$$
and obtain the resulting formula
\begin{equation}\label{eq:heat}
C = 2\pi r_{h}^{2}\frac{\left(1+9\alpha_0^2\;r^{2(1-n)}\right)}{9\alpha_0^2(1-2n)r^{2(1-n)}-1}\;.
\end{equation}
In order to analyze the thermal stability, we plot the resulting figures of heat capacity in terms of horizon as shown in Fig.(\ref{fig:first}).

\begin{figure*}
\centering
\includegraphics[width=6cm,height=6cm]{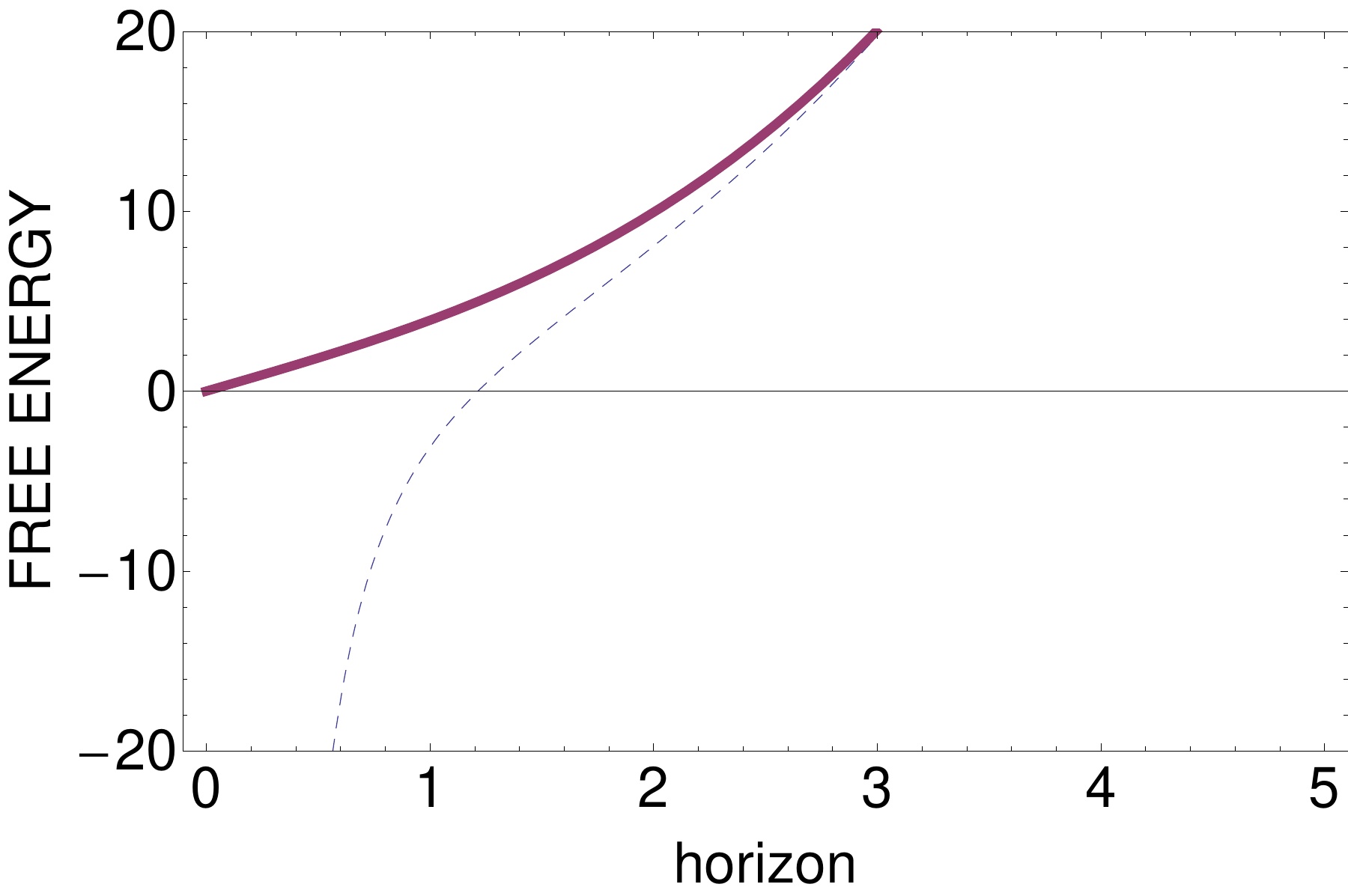}
\includegraphics[width=6cm,height=6cm]{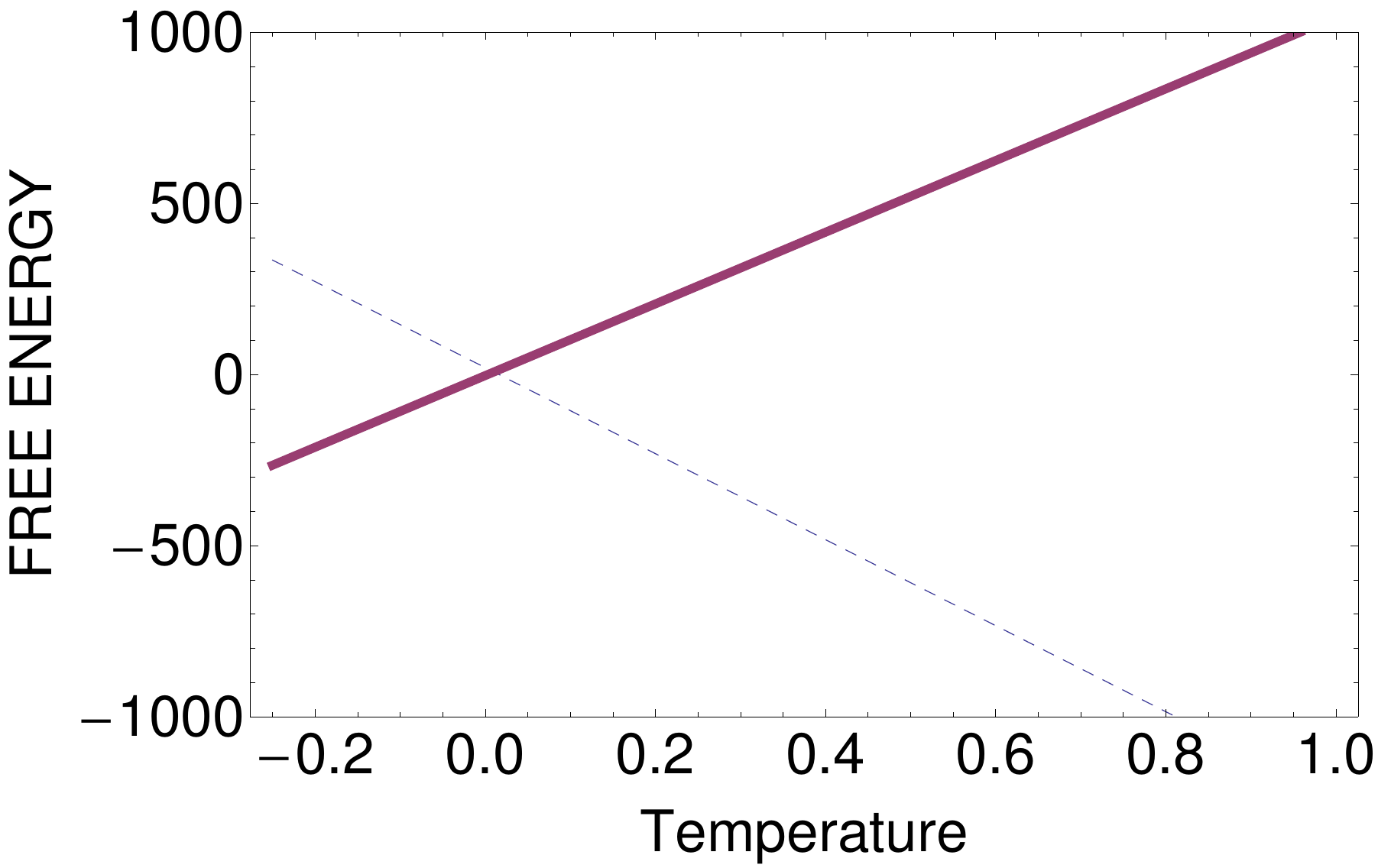}
\caption{The present figures show free energy as a function of the horizon and temperature for the allowed values of $n$. In the left panel is shown the behavior for the range with the allowed values $-1/2 \leq n \leq 1.8$ (solid line) and for the values $-1/2 \leq n \leq 3$ (dashed line) with free energy as a function of the horizon. In the right panel we compare free energy to temperature and is shown the behavior for the ranges for the allowed values $-1/2 \leq n \leq 1.8$ (solid line) and  $-1/2 \leq n \leq 3$ (dashed line).}\label{fig:second}
\end{figure*}

\begin{figure*}
\centering
\includegraphics[width=6cm,height=6cm]{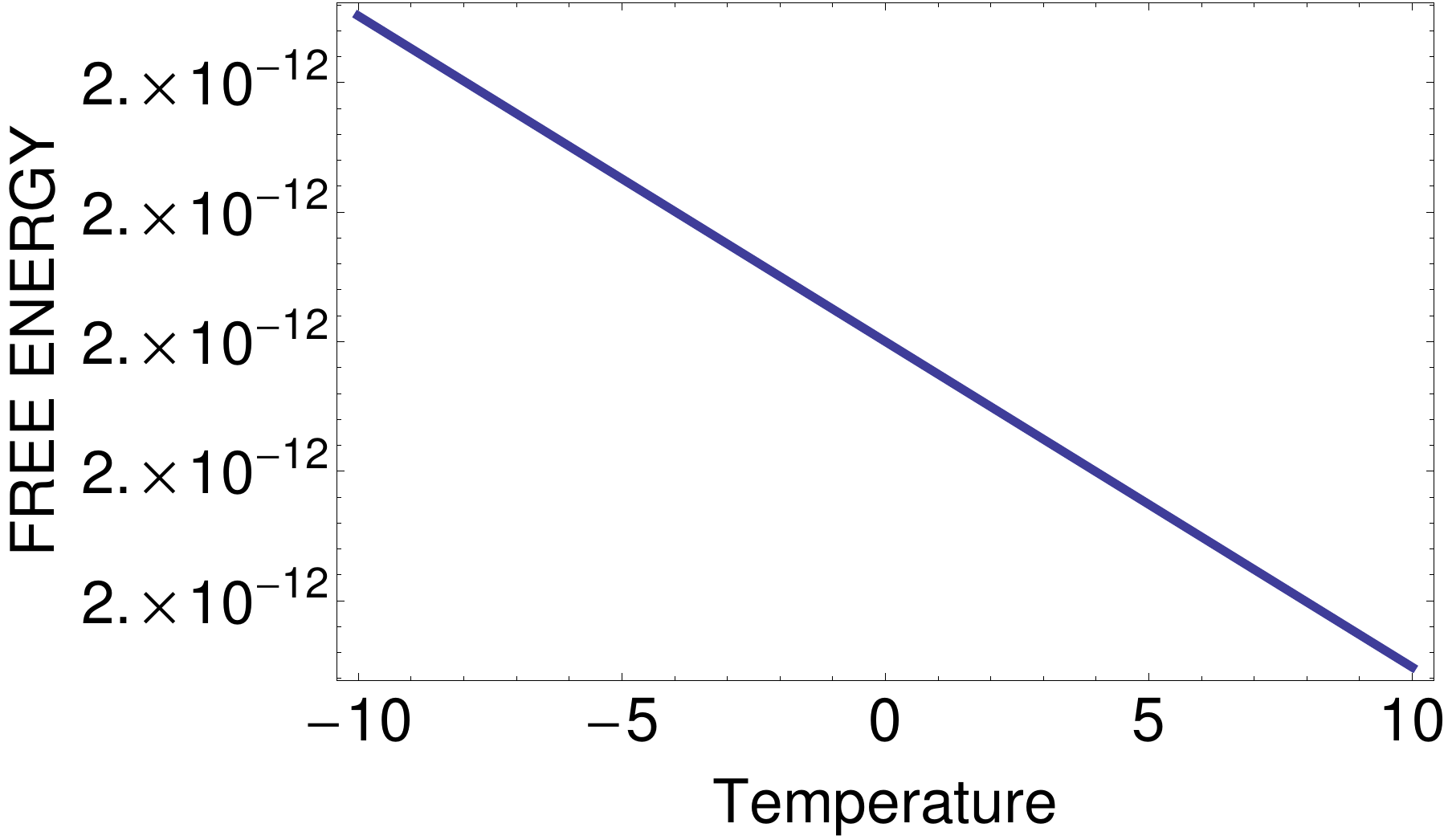}
\includegraphics[width=6cm,height=6cm]{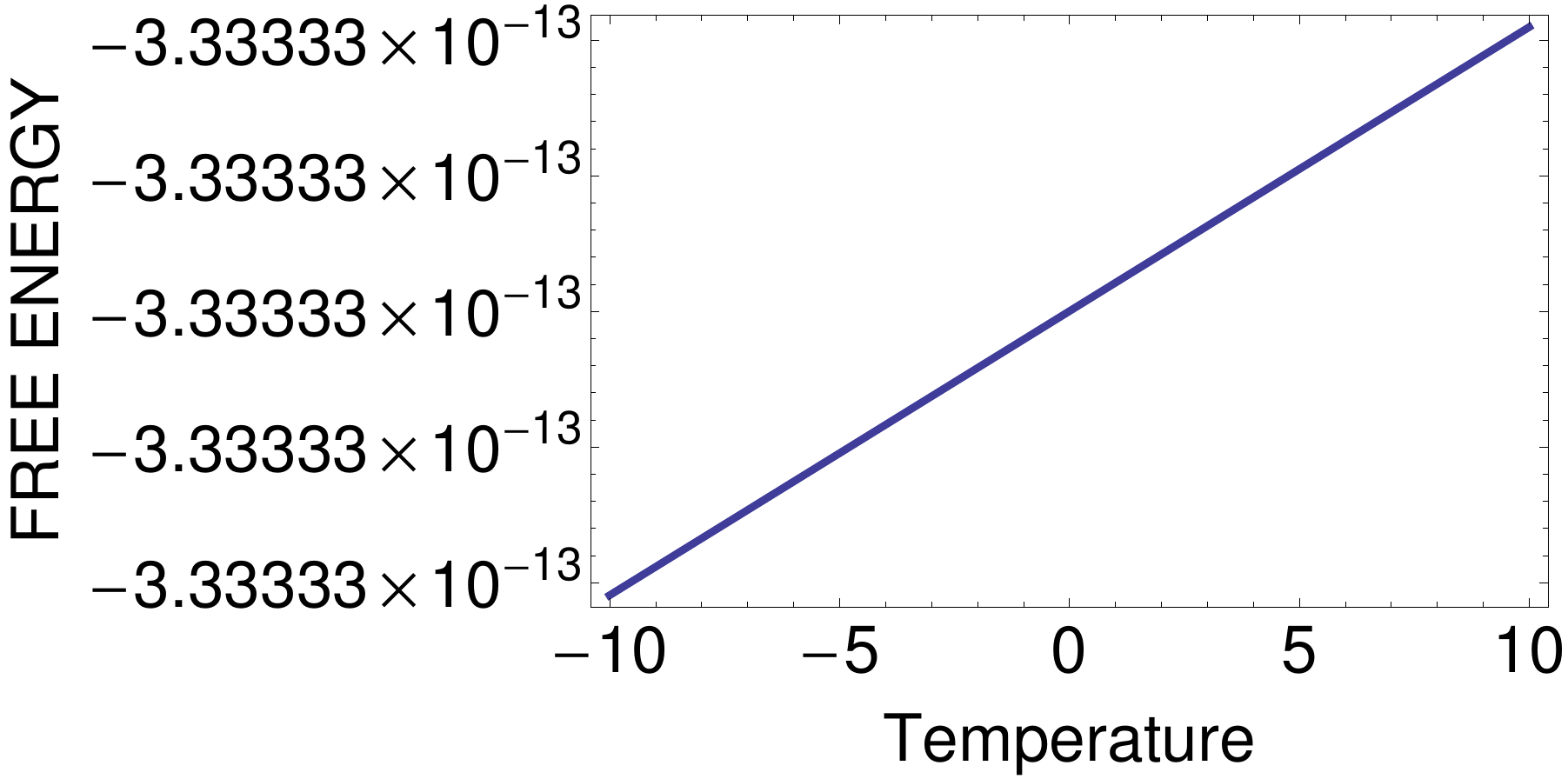}
\caption{The present figures show free energy as a function of temperature $T$ for very small black holes. In left panel is shown the behavior for the range for the allowed values $-1/2 \leq n \leq 3$ and in right panel for the case $-1/2 \leq n \leq 1.8$.}\label{fig:third}
\end{figure*}

The positive values of heat capacity indicate a local stability of the system against perturbations. As the plots indicate in the left panel Fig.(\ref{fig:first}), the initial constraints with $-1/2 \leq n \leq 3$, as shown in dashed line, lead to negative values for heat capacity, which means that we find an unstable black hole. Interestingly, a very different situation happens to a more tight range at $-1/2 \leq n \leq 1.8$, as indicated by the left panel in Fig.(\ref{fig:first}) in solid line. For small black holes we have an unstable solution. At the point of the horizon around at 0.5 we find the extreme points of heat capacity indicating a phase transition. Accordingly, after that point, the larger is the horizon the more stable is the black hole against perturbations. In the right in Fig.(\ref{fig:first}), we compare the case  $-1/2 \leq n \leq 3$ (solid line) to the standard Schwarzschild solution (dashed line). It results that the solution in the range $-1/2 \leq n \leq 3$ is more unstable than the Schwarzschild one.

Another important aspect is to study the global strength of stability which is given by the free energy. When it is related to temperature, it can be a valuable tool to study phase transition regimes. The Helmholtz free energy is given by
$$F= M - TS\;,$$
and with Eqs.(\ref{eq:mass}), (\ref{eq:hwtemp}) and (\ref{eq:entropy}), one can obtain
$$F = \frac{1}{2} \left( r_h + 9 \alpha_0^2 r_h^{3-2n}\right)\;. $$
The lesser values of the free energy indicate a global stability and the occurrence of Hawking-Page phase transition \cite{hawking,dehg}. We have checked two cases considering the free energy as a function of the horizons and latter as a function of temperature, as shown in the left and right panels in Fig.(\ref{fig:second}).

As the plots indicate in the left panel in Fig.(\ref{fig:second}), the initial constraints with $-1/2 \leq n \leq 3$, as shown in dashed line, lead to negative values for the free energy and then to unstable black holes as the horizon grows. A narrow window for transitory stability occurs around at $0.5 \leq r_h \leq 1$, which is compatible with the left panel in Fig.(\ref{fig:first}). A similar pattern occurs when considering the temperature as shown in the right panel in Fig.(\ref{fig:second}) for a fixed radius $R=10$. The transition phase occurs only at the origin with $T=0$, so we have the extremal case. For hot space, we do not have phase transitions due to the fact that the free energy drops to negative values, which leads to several issues \cite{sepangi2,ma,park}. Conversely, we have an unstable solution as the heat capacity indicates.

It is important to point out that in this case the higher is the free energy, the lower is the temperature leading to negative values, with can be discarded if we neglect unstable solutions. The same conclusion we find for very large black holes (with horizon $r_h \sim 10^{12}$). For very small black holes (with $r_h \sim 10^{-12}$), as shown in left panel in Fig.(\ref{fig:third}), the free-energy is positive and tends to a fixed value of $T=10$. Thus, the system is not thermically stable and transition phases do not occur. The conclusion is that in that range, it does not provide stable small black holes.

In the second case with $-1/2 \leq n \leq 1.8$, as shown in solid line in the left panel in Fig.(\ref{fig:second}), a phase transition occurs around zero free energy, not much different from the region presented in the left panel in Fig.(\ref{fig:third}). As we consider the temperature, as we go to $T=0$ in right panel in Fig.(\ref{fig:second}), the free energy goes to a maximum at zero which indicates a thermal equilibrium. It means that for $R=10$, and so on, we will not have a phase transition which is compatible with the heat capacity behavior. As a result, a large black hole will not have a phase transition as it grows if the temperature can be kept around zero, which defines an extremal black hole. Differently from the previous one, this case provides negative values for free energy which can give a more global stable situation, but with the price that the temperature must be negative. The same situation we have observed for very large black holes (with $r_h \sim 10^{12}$).

On the other hand, for very small black holes with a radius of the order of $10^{-2}$, we have observed that they reach stability next to $T=5$. At below radius of the order of $10^{-3}$ and so on, they stabilize at $T=10$, and the free-energy continues to drop down. For a horizon of the order of $10^{-12}$, free energy is quite small of the order of $10^{-13}$, which leads the conclusion that only small black holes can be stable at a certain range of temperature $0 \leq T \leq 10$. As an overall conclusion, the best solution occurs at $-1/2 \leq n \leq 1.8$ predicting large black holes locally stable, but not globally preferred.

\section{Remarks}
In this paper, we have discussed the local and global thermodynamic classical stability properties for a class of static black holes embedded in a five-dimensional bulk space-time. Applying Nash embedding theorem to a static spherical symmetric metric, we have found a modification induced by the extrinsic curvature. Due to the fact that this metric has a restricted embedding in five-dimensional bulk space, we have dealt with the homogeneity of Codazzi equation in Eq.(\ref{eq:BE2}) using the asymptotic condition on extrinsic curvature based on the smooth principle of Nash theorem. Hence, the metric perturbation and the embedding lead naturally to a higher dimensional structure.

We have shown that an analytical solution for embedded spherically static black holes is obtained. Analyzing the modified line element in Eq.(\ref{eq:line element1}) we have obtained the related horizons, which we have studied the behavior of a light ray near such horizons. As a result, we have found that the more extrinsic curvature is felted the more is the deformation on the space-time and the deviation is stronger than that one obtained from general relativity.

Concerning the horizons, we have got a set of scalar fields initially constrained by the potentials produced in solar scale. Moreover, the determination of thermal quantities for such black holes, we have found a tight constraint for such set of scalar fields varying at $-1/2 \leq n \leq 1.8$.  As an overall conclusion, we have obtained a restricted model of stable modified black holes (initially spherically symmetric) as compared to RS models. In this range, we have obtained a local stable behavior for large black holes with a phase transition at $r_h \sim 0.5$.

We also have shown that the global stability is not a preferred state possibly influenced by the restricted embedding in five-dimensions, which may seem not suffice for an appropriate description of the phenomena. Accordingly, in order to obtain a more general situation, both global and local stability, an embedding in a more larger bulk seems an unavoidable situation. The entropy is that one expected $\pi r^2_{h}$ that grows positively. In RS model solutions for a Schwarzschild back-hole seems to be unstable \cite{yoshino}. In addition, since our analysis is classical, mini black holes are not allowed.

Differently from RS model, we obtained  a richer general set of static Black holes, e.g, Bardeen-like black holes and Reissner-Nordstr\"{o}m-like black-holes. In this particular case, our tidal charge $\alpha_0$ is originated from the extrinsic curvature and has a new physics associated to it \cite{cap2014,GDE,gde2}, since it has a cosmological magnitude. As future prospects, a more general study of including rotation should be made concerning the canonical and grand canonical ensemble in a higher dimension, which are in due course.

\end{document}